\documentstyle[epsf]{article}

\newif\ifpreprint \preprinttrue

\def\e{\epsilon}

\def\s{\sigma}
\def\dd{\mbox{d}}

\def\R{\mbox{\scriptsize \bf R}}
\def\i{{\mbox{i}}}
\def\ii{{\mbox{\scriptsize i}}}

\def\k{{\mbox{\bf k}}}
\def\kk{{\mbox{\scriptsize \bf k}}}

\def\ds{\displaystyle}
\def\bra{\langle}
\def\ket{\rangle}

\def\bra{\langle}
\def\ket{\rangle}

\title
{
Iterative Perturbation Theory for Strongly Correlated Electron Systems with Orbital Degeneracy}
\author
{
Tetsuro Saso\\
Department of Physics, Faculty of Sciences, Saitama University,\\
 Urawa 338-8570, Japan
}

\date{\today}

\begin{document}
\sloppy
\maketitle

\vspace{2cm}
\begin{center}
Abstract
\end{center}

A new scheme of the iterative perturbation theory is proposed for the strongly correlated electron systems with orbital degeneracy.
The method is based on the modified self-energy of Yeyati, et al. which interpolates between the weak and the strong correlation limits, but a much simpler scheme is proposed which is useful in the case of the strong correlation with orbital degeneracy.
It will be also useful in the study of the electronic structures combined with the band calculations.

\ifpreprint
\vspace{2cm}
\else
\newpage
\fi
In the study of strongly correlated electron systems, the dynamical mean-field theory (DMFT)\cite{Georges96} is recognised to be a very powerful tool.
In DMFT, the lattice proplem is reduced to a single magnetic impurity embedded in an effective medium which should be determined self-consistently.
This impurity problem can be solved, e.g., by the numerial renormalization group method,\cite{Sakai94} which however needs a rather heavy numerical calculation.
Therefore, the iterative perturbation theory (IPT)\cite{Kajueter96} is used frequently.
This method adopts the modified perturbation theory (MPT),\cite{Yeyati93} which uses the second-order perturbation for the calculation of the self-energy of the correlated electrons but modifies it so as to interpolates between the weak and the strong correlation limits in the electron-hole symmetric and asymmetric cases phenomenologically.
Despite its approximate nature, the method can successfully decsribe both the metallic and the Mott-insulator states for the single-band Hubbard model.\cite{Kajueter96}

In the heavy fermion materials,\cite{Fulde88} it is of much importance to take account of the orbital degenracy.  For example, the 4f state with $J=5/2$ in the cerium compounds has six-fold degeneracy if the crystalline field splitting can be neglected.
The Fermi liquid theory is constructed for such systems by Yamada, et al.,\cite{Yamada87} and a numerical calculation via the self-consistent second-order perturbation theory (SCSOPT) has been performed by Kontani and Yamada\cite{Kontani97} for the periodic Anderson model (PAM) with the degeneracy $N=6$.
However, this calculation is applicable only to the case of weak correlation.

Yeyati, et al.\cite{Yeyati99} proposed an extension of MPT to the case of multi-orbitals and applied it to the quantum dot problem, which is equivalently described by the impurity Anderson model with the orbital degeneracy.
Their method reproduces the atomic (strong correlation) limit for any number of localised electrons, but is rather complicated even for $N=4$ (spin degeneracy times the double degeneracy of orbitals).
In the application to the heavy Fermion systems,  the Coulomb repulsion $U$ between f electrons is usually very large.  Therefore, it will be useful if one can find out a simpler interpolation scheme than Yeyati's for such a case.

In this Letter, I propose a new interpolation scheme for the self-energy by taking account of only f$^0$ and f$^1$ states of f-electrons in the ground state.  It interpolates between the weak and the strong correlation limits within the above Hilbert space.
The transition to f$^2$ state is included as an excitation.
The Fermi liquid properties are completely realised at low temperatures and low energies.  Therefore, it would be very useful if it is combined with the density of states obtained from the band calculations.

Following Kontani and Yamada\cite{Kontani97}, we start from PAM with the $N$-fold degeneracy for f-electrons written as
\begin{eqnarray}
  {\cal H} &=& \sum_{\kk\s} \e_{\kk} c_{\kk\s}^+c_{\kk\s} + \sum_{im} E_{f} f_{im}^+ f_{im}
  + U \sum_{i,\ell>m} n_{fi\ell} n_{fim} \nonumber \\
  &+&  \frac{1}{\sqrt{N_0}} \sum_{\kk\s im} \Bigl( V_{\kk\s m}e^{\ii\kk\cdot \R_i}c^+_{\kk\s}f_{im} + \mbox{H.c.} \Bigr).
\end{eqnarray}
Here, $N_0$ denotes the number of lattice sites, each of which is labeled by the index $i$, and
$E_f$, $\e_{\kk}$ and $V_{\kk\s m}$ are the f-level, the conduction electron energy and the mixing, respectively.
The indices $\ell$ and $m$ denote the $z$ component of the angular momentum of the f-states and $\s$ the spin of the conduction electrons.
For 4f electrons with the total angular momentum $J=5/2$, $V_{\kk\s m}$ is given by
\begin{equation}
  V_{\kk\s m} = \sqrt{4\pi} V \sum_\mu (-\s) \sqrt{\frac{\frac{7}{2}-m\s}{7}}\delta_{\mu,m-\frac{1}{2}\s} Y_{\ell=3}^{\mu}(\theta_\kk,\phi_\kk)
\end{equation}
if the conduction electrons are assumed to be in the plane wave states.
$V$ denotes the strength of the mixing.
In this Letter, the paramagnetic state with respect to spin and orbital is assumed.
The on-site Green's funciton for f-electrons $G_f(\e) \equiv \sum_{\kk} G_f(\k,\e)$ can be written in the form
\begin{equation}
  G_f(\e) = \frac{(N-2)/N}{\e-E_f-\Sigma_f(\e)} + \frac{1}{N_0}\sum_{\kk} \frac{2/N}{\e-E_f-\Sigma_f(\e)-\ds \frac{NV^2/2}{\e-\e_{\kk}}}.
  \label{eq:Gf}
\end{equation}
The $\k$-dependence of the self-energy is neglected by using the local approximation, which is justified in the limit of large spatial dimensions $d \rightarrow \infty$.\cite{MullerHartmann89}
When the self-energy is calculated by the second-order perturbation theory (SOPT), $\Sigma_f(\e)$ is given by $\Sigma_f(\e) = (N-1)Un_f + \Sigma_f^{(2)}(\e)$, where $\Sigma_f^{(2)}(\e)$ is calculated at zero temperature as
\begin{equation}
  \Sigma_f^{(2)}(\e) = \frac{(N-1) U^2}{4\pi^2} \int \dd\e' \dd \e'' G_f(\e') G_f(\e'') G_f(\e+\e'-\e''), \label{eq:self2}
\end{equation}
and $n_f$ denotes the number of f-electrons per site and orbital.
In SOPT, the unperturbed (Hartree-Fock) Green's function is substituted to $G_f(\e)$'s.
However, the Coulomb interaction between f-electrons is so large that SOPT is meaningless.  Therefore, the self-consistent SOPT (SCSOPT) is used in ref.\cite{Kontani97} where the self-consistently renormalised Green's function was used for $G_f(\e)$'s in eq.(\ref{eq:self2}).
It is well known, however, the density of states in SCSOPT can not reproduce the lower and upper Hubbard bands for $N=2$ when the correlation is large.\cite{MullerHartmann89}
In Fig. 1, I show my calculation of the f-electron density of states $\rho_f(\e)=-\pi^{-1}\mbox{Im}G_f(\e+\i 0^+)$ for PAM.
The shape of the density of states of the conduction band is assumed as a semicircle of the center at $E_0$ and the width $W$, which is taken as the unit of energy.
The calculation is done for $W=1$, $V=0.25$, $E_f=0.15$, $E_0=0.5$ and $U=0.3$.
A small but finite imaginary part is introduced in the denominator of the Green's function.The number of f electrons $n_f$ is obtained as 0.101 per orbital.
The mixing between the f and the conduction electrons opens the mixing gap, within which one can see the sharp peak that comes from the first term in eq.(\ref{eq:Gf}) representing the unmixed part of the $N$-fold degenerate f states.
The separation of the lower and the upper Hubbard peaks of the density of states are not seen here, which appears only when the iterative perturbation theory (IPT) is used for larger $U$ (see below).

In DMFT, the magnetic atom in the central cite can be regarded as embedded in the effective medium expressed by the cavity Green's function $\tilde{G}_f(\e)$, which is related to the on-site Green's function $G_f(\e)$ as\cite{Georges96}
\begin{equation}
  \tilde{G}_f(\e)= (G_f(\e)^{-1} + \Sigma_f(\e) )^{-1}.
  \label{eq:DMFT}
\end{equation}
IPT calculates the self-energy of an f-electron due to the Coulomb interaction on this impurity site by SOPT using $\tilde{G}_f(\e)$ as an unperturbed Green's function.
This works well only for the symmetric case, where even the Mott transition is properly described.\cite{Kajueter96}
In the asymmetric case, however, it fails to reproduce the correct atomic limit.
Yeyati, et al.\cite{Yeyati93} proposed to modify the second-order self-energy in a phenomenological manner, by which it becomes possible to interpolate between the weak and the strong correlation limits.
When $N=2$, the following form is assumed\cite{Kajueter96},
\begin{equation}
  \Sigma_f(\e) = U n_f + \frac{A \Sigma_f^{(2)}(\e)}{1-B \Sigma_f^{(2)}(\e)}. \label{eq:self}
\end{equation}
The parameters $A$ and $B$ are determined so as to reproduce the correct $|\e| \rightarrow \infty$ and $U \rightarrow \infty$ limits, respectively:
\begin{equation}
  A = \frac{n_f(1-n_f)}{\tilde{n}_f(1-\tilde{n}_f)},
\end{equation}
\begin{equation}
  B = \frac{U(1-n_f) + E_f - \tilde{E}_f}{U^2\tilde{n}_f(1-\tilde{n}_f)},
\end{equation}
where $n_f=n[G_f]$ and $\tilde{n}_f=n[\tilde{G}_f]$.  The functional $n[G]$ means the electron number calculated by
\begin{equation}
  n[G] = \int \dd\e f(\e) \left(-\frac{1}{\pi}\right) \mbox{Im} G(\e+\i 0^+).
  \label{eq:n}
\end{equation}
$\tilde{E}_f=E_f+Un_f$ denotes the Hartree-Fock level.
We call this scheme as the modified perturbation theory (MPT).
Recently, I discussed its microscopic basis in \cite{Saso00}.

Since the modified form of the self-energy is not conserving, it does not satisfy the Luttinger sum rule\cite{Luttinger60,Ohkawa84}:
\begin{equation}
  n_f+n_c=\frac{1}{N_0}\sum_{\kk} \theta(\mu-E_{\kk-}).
\end{equation}
Here $\mu$ ($=0$) denotes the chemical potential and $\theta(x)$ denotes the step function.
$E_{\kk-}$ denotes the lower branch of the quasi-particle band and I have assumed the case that $\mu$ lies in the lower band.
$n_c$ denotes the number of conduction electrons.
It is possible to approximately satisfy the sum rule by subtracting the zero-energy part as\cite{Yamada79}
\begin{equation}
  \Sigma_f'(\e) = \Sigma_f(\e) - \Sigma_f(0).
  \label{eq:sub}
\end{equation}
We follow this prescription in the present Letter.
If the sum rule must be satisfied explicitly, one can regard $\tilde{E}_f$ as a free parameter and determine it by the Luttinger sum rule.\cite{Yeyati93,Takagi99,Saso99}

For $N>2$, the f-electron Green's function in the atomic limit is given by\cite{Yeyati99}
\begin{eqnarray}
  G_m^a(\e) &=& \frac{\bra\prod_{\ell(\neq m)} (1-n_\ell)\ket}{\e-E_f}
  + \sum_{\ell\neq m}\frac{\bra n_\ell \prod_{s(\neq\ell\neq m)} (1-n_s)\ket}{\e-E_f-U} + \cdots \nonumber \\
 & & \hskip 2cm + \frac{\bra\prod_{\ell(\neq m)} n_\ell\ket}{\e-E_f-(N-1)U},
\end{eqnarray}
where the suffices $m, \ell, s, \cdots$ label the orbital states of f-electrons.
Yeyati, et al.\cite{Yeyati99} found a scheme which interpolates between this $G_m^a(\e)$ and the second-order perturbation theory.
Their formula, however, needs the average values of the products of the f-electron number operators in the different orbitals, $\bra n_m n_\ell n_s \cdots \ket$, which must be determined self-consistently.
The procedure is complicated enough even for $N=4$.

In the realistic heavy Fermion materials, it is quite often the case that $U$ is large enough compared to the resonance width of the f-states.
Then the ground state is restricted to f$^0$ and f$^1$ states in the case of Ce, so that the above Green's function can be approximated by
\begin{equation}
  G_m^a(\e) \simeq \frac{1-\sum_{\ell(\neq m)}\bra n_\ell\ket}{\e-E_f} + \frac{\sum_{\ell(\neq m)}\bra n_\ell\ket}{\e-E_f-U},
  \label{eq:Gfa}
\end{equation}
since the terms which include more than two $n_\ell$'s vanish.
The first and the second terms in eq.(\ref{eq:Gfa}) represent the transitions f$^1 \rightarrow$ f$^0$ and f$^1 \rightarrow$ f$^2$, respectively.

In the same manner as $N=2$, we start from the SOPT self-energy, eq.(\ref{eq:self2}),
and modify it into the form of eq.(\ref{eq:self}) so as to interpolate between the weak and the strong correlation limits.
Noting that $\sum_{\ell(\neq m)}\bra n_\ell\ket=(N-1)n_f$, we can easily determine the parameters $A$ and $B$ as
\begin{equation}
  A = \frac{n_f[1-(N-1)n_f]}{\tilde{n}_f(1-\tilde{n}_f)},
  \label{eq:A}
\end{equation}
\begin{equation}
  B = \frac{1-2\tilde{n}_f}{U(N-1)\tilde{n}_f(1-\tilde{n}_f)},
  \label{eq:B}
\end{equation}
by the condition that $\Sigma_f(\e)$ reproduces the correct high energy limit and the atomic limit, respectively.
(We have used the relation $\tilde{E}_f-E_f=U(N-1)n_f$.)

The procedure proposed in this Letter can be summarised as follows.  We start from the Hartree-Fock solution to $G_f(\e)$ and calculate the second-order self-energy eq.(\ref{eq:self2}).
The Fourier transformation method\cite{MullerHartmann89} is used in the calculation.
We then construct the modified self-energy by eq.(\ref{eq:self}) with the Hartree-Fock term $(N-1)Un_f$ instead of $Un_f$. $A$ and $B$ are given by the above equations (\ref{eq:A}) and (\ref{eq:B}), and
$\tilde{n}_f=n_f$ at this stage.
To fulfil the Luttinger sum rule approximately, we subtract the constant term from the self-energy as eq.(\ref{eq:sub}).
Next, we reconstruct $G_f(\e)$ by eq.(\ref{eq:Gf}) using this modified self-energy, and calculate the cavity Green's function $\tilde{G}_f(\e)$ by eq.(\ref{eq:DMFT}).
The self-energy eq.(\ref{eq:self2}) is calculated again by using $\tilde{G}_f(\e)$, and repeat the above cicle until the convergence of the solution is achieved. $n_f$ and $\tilde{n}_f$ are recalculated at each iteration.

As a numerical example, we show in Fig. 2 the f-electron density of states for $N=6$, $W=1$, $V=0.2$, $E_0=0$, $U=1$ and (a) $E_f=-0.3$ and (b) $E_f=-0.5$.  The insets enlarge the low energy regions.  The case (a) is metallic, while the case (b) is just on the border between the metallic and the (Kondo) insulating states.
The number of f-electrons is (a) $n_f=0.102$ and (b) $n_f=0.127$, both of which satisfy the Luttinger sum rule within 3\% of accuracy.
Because of the strong correlation, the Kondo peak is renormalised to a small width and is further split by the mixing.
One can see the large peak above the chemical potential, which represents the spectra for the transition f$^1 \rightarrow$ f$^2$ and has a large weight due to the degeneracy.
This peak and the peak below the chemical potential corresponding to the f$^1 \rightarrow$ f$^0$ transition is absent in the SCSOPT calculation shown in Fig. 1 and can not be reproduced in the stronger correlation within that scheme.

The new scheme proposed in this Letter can describe the electronic states of the strongly correlated electron systems with orbital degeneracy in a manner much simpler than that by Yeyati, et al.\cite{Yeyati99}  It can be easily combined with the density of states obtained from the band calculations, and will be used for an analysis of the realistic models of heavy Fermion materials with the strong correlation.

The present scheme can be applied only to zero temperature when the Luttinger sum rule is forced to be fulfilled strictly.
Since it is fulfilled only approximately in the present calculation, it is possible to apply the present scheme to finite temperatures to obtain semiquantitative but reasonable results.\cite{Takagi99}

Corrections of the next leading order of the $1/d$-expansion can be taken into account as a spin-fluctuation by using the technique proposed in \cite{Saso99}.
Extension to the case with the crystalline field splitting, and calculations of the magnetic susceptibility and other physical quantities will be the important issues and are now under study.

This work is supported by Grant-in-Aid for Scientific Research No.11640367
from the Ministry of Education, Science, Sports and Culture of Japan.

\ifpreprint
\vspace{2cm}
\else
\newpage
\fi
\begin{figure}[h]
\epsfxsize=10cm
\centerline{\epsfbox{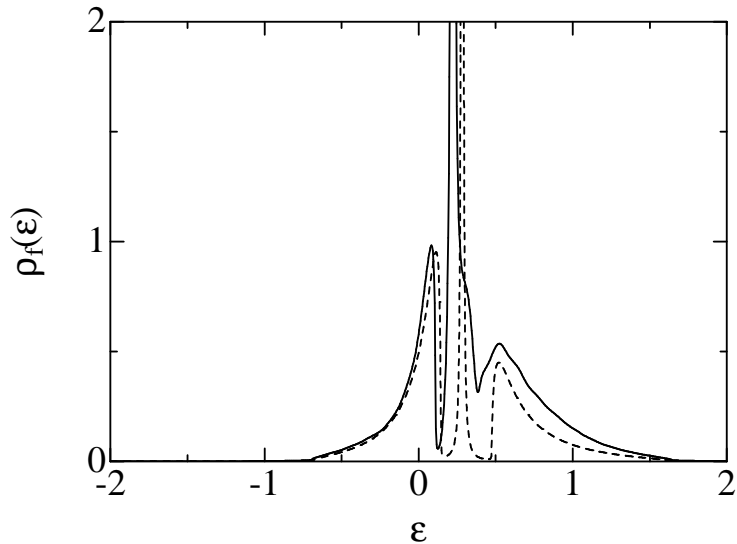}}
\caption{The f-electron density of states for $N=6$, $W=1$ (the unit of energy), $V=0.25$, $E_f=0.15$ and $U=0.3$ is shown.  The straight and dashed lines indicate the SCSOPT and the Hartree-Fock results, respectively.}
\label{fig1}
\end{figure}
\begin{figure}
\epsfxsize=10cm
\centerline{\epsfbox{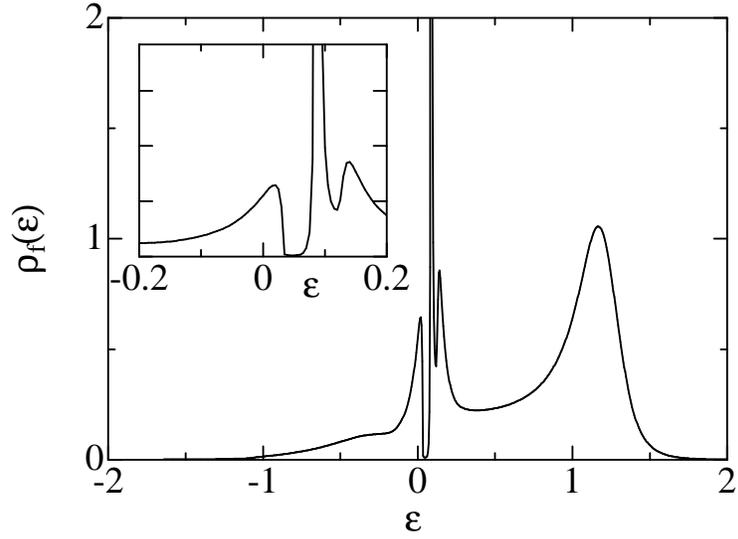}}
\vspace{2cm}
\epsfxsize=10cm
\centerline{\epsfbox{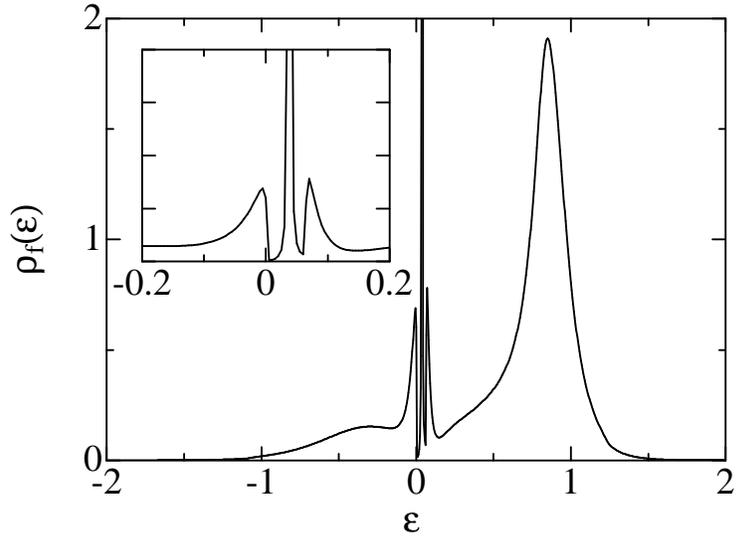}}
\vspace{1cm}
\caption{The f-electron densities of states for $N=6$, $W=1$, $V=0.2$, $U=1$ and (a) $E_f=-0.3$ and (b) $E_f=-0.5$ are shown.  The insets display the low energy parts.}
\label{fig2}
\end{figure}

\end{document}